\documentclass{llncs}
\usepackage{makeidx}  
\usepackage{graphicx} 
\usepackage{subeqnar} 
\usepackage{multicol} 

\usepackage{amsmath}
\usepackage{amssymb}
\usepackage{amscd}
\usepackage{latexsym}
\usepackage{epsfig}

\newtheorem{defi}{Definition}

\newtheorem{Theorem}[defi]{Theorem}

\date{}

\begin{document}


\title{Three lines proof of the lower bound for the matrix rigidity}
\titlerunning{Three lines proof of the lower bound for the matrix rigidity}

\author{Gatis Midrij\= anis
}
\authorrunning{Gatis Midrij\= anis }

\institute{University of California, Berkeley. \\
\texttt{gatis@berkeley.edu}}

\maketitle

\begin{abstract}
The rigidity of a matrix describes the minimal number of entries one
has to change to reduce matrix's rank to $r$. We give very simple
combinatorial proof of the lower bound for the rigidity of Sylvester
(special case of Hadamard) matrix that matches the best known result
by de Wolf(2005) for Hadamard matrices proved by quantum information
theoretical arguments.
\end{abstract}

\section{Introduction}\label{sec:Intro}
\subsection{Problem} \label{sec:prob}

Changing some entries of a complex matrix can reduce its rank. The
rigidity of a matrix $M$ is the function $R_M(r)$, which for a given
rank $r$, gives the minimum number of entries of $M$ which one has
to change in order to reduce $M$'s rank to $r$ or less. More
formally,
$$R_M(r) = \min\limits_{rank(\widetilde{M})\leq r}\{weight(M-\widetilde{M})\},$$
where $weight$ denotes the number of non-zero entries. In other
words, large rigidity shows that the matrix's rank is stable under
perturbations. It is easy to see that $R_M(r) \geq n-r$ for any full
rank matrix $M$, because change of one entry reduces the rank by at
most 1.

\subsection{History} \label{sec:hist}
In this section I survey all known results for the matrix rigidity
over infinite fields up to my best knowledge. There have been done a
lot of work on rigidity on finite fields \cite{Friedman} and on
restricted and generalized versions of the rigidity problem as well
\cite{PRS}.

The rigidity was defined by Valiant~\cite{ValiantGr,ValiantT}; a
similar notion independently was proposed by Grigoriev~\cite{Grig}.
The main motivation to study rigidity is that good lower bounds on
rigidity would give important complexity results in other
computational models, like linear algebraic circuits and
communication complexity. For communication complexity
$(0,1)$-matrices are especially important. Valiant showed $R_M(r)
\geq (n-r)^2$ for "almost all" matrices $M$. Pudlak and
Rodl\cite{PR} showed a similar result for $(0,1)$-matrices. However,
to show a good lower bound for an \textbf{explicit} matrix still
remains unsolved task.

The most interesting matrix probably is Hadamard matrix. Pudlak and
Savicky \cite{PS} showed that for any Hadamard matrix $H$, $R_H(r)
=\Omega(\frac{n^2}{r^4 \log^2 r})$, Razborov~\cite{RazUnp} improved
their result to $R_H(r) =\Omega(\frac{n^2}{r^3 \log r})$.
Grigoriev\cite{GrigUnp} and Nisan~\cite{Nisan} independently
observed an easy method to get lower bound for any \emph{totally
non-singular} matrix $M$ (i.e. a matrix in which all submatrices are
non-singular) $R_M(r) = \Omega(\frac{n^2}{r})$. Similar strategy was
used by Alon~\cite{Alon} to improve rigidity of Hadamard matrix
$R_H(r) = \Omega(\frac{n^2}{r^2})$, Lokam\cite{LokamSp} to give an
alternate proof of the same result, Kashin and Razborov \cite{KR} to
prove $R_H(r) \geq \frac{n^2}{256r}$ and de Wolf\cite{Wolf} to use
quantum information theoretical arguments to give a neat proof of
$R_H(r) \geq \frac{n^2}{4r}$ (the last result holds for any
orthogonal matrix where all entries have the same magnitude,
including Discrete Fourier Transform).

There are results for other matrices as well. Razborov~\cite{RazRus}
showed $R_M(r) = \Omega(\frac{n^2}{r})$ if $M$ is the generalized
Fourier Transform matrix or the inverse of the Vandermonde matrix.
Kimmel and Settle\cite{KS} gave the lower of the rigidity of the
triangular matrix $T$, their result in simplified form looks like
$R_T(r) \approx \Omega(\frac{n^2}{r})$. Independently, Pudlak and
Vavrin \cite{PV} determined the exact value of $T$, particulary, for
a large $n$ but small $r$ it is like $R_T(r) \approx
\frac{n^2}{4r}$. Pudlak\cite{Pudlak} showed that $R_M(r) =
\Omega(\frac{n^2}{r})$ if $M$ belongs to a class of matrices called
Densely Regular, that includes triangular matrix, Vandermonde
matrices, shifters and parity shifters. Shokrollahi et al.
\cite{SSS} showed that $R_C(r) = \Omega(\frac{n^2}{r}\log
\frac{n}{r})$ for a Cauchy matrix $C$. Codenotti et al.\cite{CPR}
studied the rigidity of some matrices under combinatorial
assumptions.

Lokam\cite{LokamVan,LokamUnp} gives some quadratic lower bounds for
"less explicit" matrices. Landsberg et al.\cite{LTV} gave
geometrical interpretation of matrix rigidity.

However, these results do not give a superlinear rigidity for an
explicit matrix when $r=O(n)$. Lokam\cite{LokamVan} observes that a
method used in all those results (and this paper as well) by getting
"candidate" matrices that are close to full rank does not give a
results like $R(r) = \omega(\frac{n^2}{r}\log \frac{n}{r})$.

Codenotti~\cite{Cod} gives a survey paper on the matrix rigidity
problem as well as some interesting problems.

In this paper we give a simple proof in "three lines" of $R_S(r)
\geq \frac{n^2}{4r}$ for any Sylvester matrix $S$ (special case of
Hadamard matrix). The same proof works for other "well behaved"
matrices, like Discrete Fourier Transform. However, our main
contribution is the simplicity of the proof.

\subsection{Matrices} \label{sec:matr}

If $A=(a_{ij})$ and $B=(b_{kl})$ are matrices of size $m \times n$
and $p \times q$ respectively, the \emph{Kronecker product} $A
\otimes B$ is the $mp \times nq$ matrix made up of $p \times q$
blocks, where the $(k,l)$ block is $b_{kl}A$.

\emph{Sylvester matrix} $S(n)$ of order $n:=2^k$ is $n \times n$
matrix made by iterating Kronecker product of $k$ copies of the
following matrix

$$ S(2) = \left(
  \begin{array}{cc}
    + & + \\
    + & - \\
  \end{array}
 \right)$$
where $+$ and $-$ denotes $+1$ and $-1$ respectively.

For example, $$ S(4) = \left(
  \begin{array}{cccc}
    + & + & + & + \\
    + & - & + & - \\
    + & + & - & - \\
    + & - & - & + \\
  \end{array}
\right)$$

Sylvester matrices are special case of Hadamard matrices. A real
valued matrix $H$ is called \emph{Hadamard matrix} iff $HH^T=nI$.

\emph{Discrete Fourier Transform} is $n \times n$ matrix $FN(n) =
(f_{jk})$ defined by $f_{jk}:= \omega^{(j-1)(k-1)}$, where $\omega
:= e^{\frac{2 \pi i}{n}}$ and $i:= \sqrt{-1}$.

\section{Proof}
\label{sec:proof}

\begin{Theorem}
If $S(n)$ is a Sylvester matrix and $r \leq n/2$ is a power of $2$
then
$$R_{S(n)}(r) \geq \frac{n^2}{4r}.$$
\end{Theorem}
In other words, for any $n \times n$ matrix $\widetilde{S}$ such
that $rank(\widetilde{S}) \leq r$ holds
$$weight(S(n)-\widetilde{S}) \geq \frac{n^2}{4r}.$$

\proof Assume the opposite, $weight(S(n)-\widetilde{S}) <
\frac{n^2}{4r}$. Let uniformly divide $\widetilde{S}$ in
$(\frac{n}{2r})^2$ submatrices $\widetilde{S}_{ij}$ of size $2r
\times 2r$. By a counting argument, there exists $i,j$ s.t.
$weight(S(2r)-\widetilde{S}_{ij}) < r$. Thus, $rank(\widetilde{S})
\geq rank(\widetilde{S}_{ij}) > 2r-r=r$. Contradiction. \qed

The same proof works for $R_{FT(n)}(r) \geq \frac{n^2}{4r}$, where
$FT(n)$ denotes $n \times n$ Discrete Fourier Transform matrix,
because DFT matrix where columns with even indexes are written first
is represented as a matrix
$$ \left(
  \begin{array}{cc}
    FT(n/2) & \ \  \omega^j FT(n/2) \\
    FT(n/2) & \ \  -\omega^{j-n/2} FT(n/2) \\
  \end{array}
 \right)$$
where $j$ denote the index of a row. Since rows of a $FT(n)$ are
orthogonal and multiplication by some constant does not change this
property, each submatrix can be recursively divided again and again,
by getting full rank submatrices. This is the only property of
matrices we need in the proof.

\end{document}